\documentclass[a4paper,12pt,twoside]{article}
\usepackage{amssymb}
\usepackage{graphicx}
\usepackage{amsmath}
\usepackage{amsthm}
\usepackage{epstopdf}
\usepackage{setspace}
\usepackage[hmargin=1.in,vmargin={.8in,1.2in},centering]{geometry}
\usepackage[belowskip=10pt]{caption}
\usepackage{natbib}
\usepackage{bm}
\usepackage{comment}
\usepackage{hyperref}
\hypersetup{linkcolor=blue,citecolor=blue,urlcolor=blue,colorlinks=true}
   \usepackage{rotating}
   \usepackage{longtable}
   \usepackage{float}
   \usepackage{subcaption}
   \usepackage{tabularx} 
   \usepackage{booktabs}

\usepackage{lettrine} 
\usepackage{paralist} 
\usepackage{fancyhdr} 
\usepackage{titlesec} 
\usepackage{appendix} 
\usepackage[bottom]{footmisc} 
\usepackage{libertine}
\usepackage[libertine]{newtxmath}   

\DeclareGraphicsExtensions{.eps}

\ifx\pdfoutput\undefined
\else
   \fi
   
\begin{document}

\title{Corporate Finance in the Age of Fintech: Scenarios and Challenges}

\author{ \vspace{1cm} 
	Nicola Borri \thanks{Lian Group Chair in Fintech and Blockchain Technology, Department of Economics and Finance, LUISS University, Rome 00197, Italy. Contact email: \href{nborri@luiss.it}{nborri@luiss.it}; web: \href{https://www.nicolaborri.com}{www.nicolaborri.com}. The author  acknowledges the discussions on the topics of this paper provided by Yukun Liu, Fiorenzo Manganiello, Kirill Shakhnov, and Aleh Tsyvinski.}
}

\date{\today
}
\maketitle

\vspace{-0.5cm}
\begin{center}
\begin{abstract}
\noindent Blockchain is a technological innovation that has the potential to radically change our financial markets by providing an alternative management approach to the ``promise market'', which is the foundation of our financial systems. Its disruptive potential also extends to corporate finance, where blockchain is beginning to influence valuation methods and capital allocation strategies, offering new perspectives on how companies are assessed and financed. However, for a new financial architecture based on blockchain and advancements in technology---what is commonly referred to as Fintech---to replace, in whole or in part, traditional finance, it will need to overcome significant challenges such as regulation, environmental sustainability, its association with illegal activities, and achieving greater efficiency in cryptocurrency markets. For this reason, the future of Fintech is likely to be more conventional---yet also more transparent, efficient, and regulated---ultimately evolving to resemble the traditional finance we know.
\end{abstract}
\end{center}



\thispagestyle{empty}
\clearpage

\newpage

\section{Introduction}\label{sec:intro}

The objective of this article is to discuss the scenarios and challenges for the future of finance, commonly referred to as Fintech and generally, though not exclusively, linked to one of the most significant technological innovations in recent years: blockchain, with its decentralized information validation mechanism.

To understand the potential role of blockchain—and consequently of Fintech—in the near future, it is beneficial to start by addressing a fundamental and basic question: ``What is a financial asset or financial instrument?" A possible answer, paraphrasing Wikipedia's definition, might simply be: ``A financial instrument is a non-physical asset whose value depends on obligations and conditions defined in a contract, such as a bank deposit account, a bond, or company shares."\footnote{For further details, see \href{https://en.wikipedia.org/wiki/Financial_instrument}{https://en.wikipedia.org/wiki/Financial\_instrument}.}

A more precise definition would identify a financial instrument as a ``right to receive something in certain states of the world." This essentially aligns with Arrow and Debreu’s definition for Arrow-Debreu securities (see, \cite{arrow1954existence}, or chapter 3 in \cite{cochrane2009asset}), whose payoffs are contingent on specific states of the world. For instance, an option contract allows one to buy or sell a particular financial instrument at a given price on a certain future date. Similarly, a bond guarantees repayment at maturity, provided the debtor is not in default. Fundamentally, any financial asset can be described as a combination (or portfolio) of Arrow-Debreu securities.

I would like to propose an complementary interpretation of financial instruments based on the existence of a ``market for promises" (see, \cite{levine_crypto} and \cite{zhang_promises}). According to this definition, financial instruments are fundamentally promises. Each financial instrument represents a promise to deliver something to someone, typically conditional upon uncertain future events. This perspective emphasizes the probability distribution of possible states of the world, foundational to Arrow-Debreu securities and their payoffs.

For example, consider a promise to pay a certain amount based on the level of the S\&P 500 index at a future date; this promise is essentially a futures contract. Another example is a promise to pay a share of a company's profits indefinitely, equivalent to an equity share. Similarly, a promise to make regular payments, with the creditor allowed to seize collateral (like a house or car) in case of default, mirrors a mortgage or collateralized debt contract. Thus, we can conceive the entire financial system as a market exchanging promises.

Traditionally, the judicial and legal system safeguards the market for promises. Indeed, one of the historical reasons behind the emergence of geographical entities administered by a central authority (and, more generally, the development of the rule of law) is precisely the necessity of having a mechanism that could ensure the proper functioning of the market for promises, and thus of a financial system capable of supporting economic activity and growth (see, e.g., \cite{harari2014sapiens}, or \cite{umbeck1977california}'s account of the emergence of property rights during the California gold rush). The importance of property rights and their protection for economic growth similarly arises from their crucial role in maintaining a well-functioning market for promises.

Recently, a technological innovation has led to the creation of the blockchain, opening new scenarios for financial markets that were unimaginable just a few years ago. We can think of the blockchain as an enormous database maintained in a decentralized manner and secured through cryptography and verification by millions of validators, called ``miners." Thus, with the blockchain, it is the users themselves who, in a decentralized way, safeguard the market for promises. This technological innovation lies at the foundation of Fintech and, perhaps for the first time in history, surpasses the role of the judicial and legal systems as the fundamental pillars of the market for promises.

Classical economic theory does not explicitly consider the market for promises. Instead, it typically analyzes simpler markets. Take, for example, the market for potatoes.\footnote{This example is from \cite{zhang_promises}.} When I wish to buy potatoes, I go to the market with cash, hand the money to the seller, and receive potatoes in exchange. Once the money is exchanged for potatoes, the transaction is complete. There is no need to interact again with the seller for the same transaction in the future (although I might return later to buy more potatoes!). For this reason, buying potatoes at the market does not require a particularly high level of ``trust" for the transaction to be completed successfully. In contrast, the market for promises requires a much higher level of trust. Consider, for example, a ``bet" (or investment) that the S\&P500 index will decline. This bet is equivalent to a futures contract. In this case, trust is fundamental: the investor must trust the counterparty to honor their obligation and pay their debt if they lose. In technical terms, this risk is commonly defined as counterparty risk, and modern financial markets rely on intermediaries, such as clearing houses, to curb this risk. The market for promises can function only when we are convinced that promises will be kept (at least with a probability greater than zero).

This type of market characterizes many typical and complex contracts found in financial markets. If I start a new company and seek to raise capital, or equity, investors must trust that the company's management can generate profits to be distributed to them. When a bank lends money to a family to purchase a house, it must know that it can seize the collateral (typically through a mortgage on the house) if the borrower fails to make the required loan payments.

The market for promises---that is, the modern financial system---can only exist when founded upon a system capable of enforcing promises among private subjects, individuals, and/or companies. For this reason, the modern financial system relies on the proper functioning of the judicial and legal systems. The state holds a monopoly on the use of ``violence," and one of the applications of this monopoly power is precisely contract enforcement. Finance, in the sense of a financial system, would not be possible unless these promises were upheld, and these promises remain credible only as long as a government, or the judicial system, has the willingness and ability to enforce them, or to impose the consequences stipulated for those who breach them.

In this sense, it is not hyperbole to say that finance today is fundamentally a derivative of the legal system designed to enforce promises. This simple observation explains why finance is concentrated in developed countries, particularly those with better-functioning legal systems where property rights are more respected. Conversely, in countries with weaker judicial systems, financial markets are also less developed; for example, credit is often provided within family groups rather than through banks, since informational asymmetries (such as the ability to repay debts) and moral hazard (such as the willingness to repay debts) tend to be lower within these groups (see, e.g., \cite{porta1998law} and \cite{putnam1994making}).

The state's monopoly on justice can be a double-edged sword. Indeed, it implies that the enforcement and control of promises between economic agents depend on the decisions and, for example, political moods of a government. The goal of governments is not always to guarantee these promises: governments aim to maintain power and can change their minds regarding previously made commitments. A classic example is governments in developing countries, which often expropriate private assets. However, even in developed countries, it is not difficult to find examples of governments failing to uphold their promises, even in recent times. For instance, during the Covid-19 pandemic, governments in many developed countries suspended payment obligations for debtors, delayed mortgage installments, and extended lease contracts, among other measures. Certainly, an event of the magnitude of the pandemic, fortunately, does not occur frequently. Yet, even during the great financial crisis of 2009, we witnessed partially similar situations in developed countries, with public interventions radically altering previously made promises (such as debt moratoriums and rescue interventions for struggling financial intermediaries). Finally, the recent unexpected surge in inflation can also be viewed as another instance of governments failing to uphold their promises---in this case, the commitment to safeguarding the purchasing power of investors' savings.

In general, governments have a reputational incentive to uphold their promises. When they do so, businesses and individuals place their trust in the government. This trust is essential for undertaking economic initiatives that enable economic growth. However, the financial system in each country is so large that it often becomes difficult, if not impossible, for a government to guarantee all the promises it has made. In these circumstances, the government's role as an impartial arbiter inevitably becomes compromised, as diverse political objectives may influence its choices and decisions. This observation, tied to human nature itself, constrains the type and quantity of promises that can be made, and consequently, the scale of economic activity.

Blockchains are fundamentally alternative systems for guaranteeing and enforcing promises, radically different from those known until today in our history. In blockchain-based systems, promises are enforced and guaranteed by miners who, in a reasonably competitive mining market, have no incentive other than executing promises made between individuals or economic entities in exchange for compensation, the so-called ``gas fees.'' (see, e.g., \cite{abadi2018blockchain}, \cite{catalini2020some}, \cite{leshno2020bitcoin}, and \cite{cong2021decentralized}). In this sense, blockchain can be seen as a technological innovation that has created a new and universal system for effectively enforcing promises with a minimal level of discretion.

For instance, today there exist automatic market-making (AMM) protocols, such as Uniswap. A market maker is a financial institution providing liquidity to markets by buying and selling securities. In traditional financial systems, banks or financial intermediaries typically serve as market makers. Uniswap, however, is an automated market maker, meaning it is a protocol allowing any investor to participate in market-making activities in exchange for compensation. The terms of the contract between Uniswap and users who supply liquidity to the market are not enforced by the judicial system, but rather by a smart contract---a piece of code written in Solidity---executed by miners on the Ethereum blockchain.\footnote{Solidity is a programming language used for writing smart contracts that run on various blockchain platforms, notably Ethereum. A smart contract is a piece of code designed to automatically execute operations based on the terms (conditions) of a contract. The Ethereum blockchain is specifically designed to execute smart contracts.} Although Uniswap is currently a platform exclusively for trading cryptocurrencies, it is not difficult to imagine a near future where this same technology could be utilized to build markets for trading traditional financial instruments, such as stocks, bonds, and currencies, thereby competing with traditional platforms (see, e.g., \cite{cong2021tokenomics}).

Another example of blockchain-related financial innovation is represented by lending protocols at the core of frontier of Fintech, such as Aave, which allow anyone to borrow funds by offering risky assets, like bitcoin, as collateral. Aave, fully automatically and without discretion, is capable of evaluating the value of the collateral provided. Furthermore, Aave is also capable, again automatically, of seizing the collateral or liquidating part of it if the borrower becomes insolvent. Aave functions similarly to a bank or an institution offering margin lending to an investor. The primary difference is that Aave is a bot---a program or piece of code---that cannot be modified once registered on the blockchain. Hence, Aave addresses a problem already dealt with by traditional finance by providing, in effect, a similar product but utilizing a different technology for its implementation.\footnote{For the effect of Fintech entry in the lending market, the competition with traditional financial institutions and the welfare consequences, see \cite{vives2025fintech}.}

These two examples, Uniswap and Aave, illustrate how a large number, perhaps even all, of the financial services offered by traditional financial systems can be provided by ``programs'' written in languages such as Solidity and registered on a blockchain. In this sense, Fintech has the theoretical potential to replace traditional finance. For the first time in human history, we have a mechanism for guaranteeing promises in the market that is different from the one based on government authority and the monopoly over the use of force, thus laying new foundations upon which the financial system can rest.

What will the implications of all this be for our way of life? It is likely that, at least initially, these implications will be limited in developed countries. In fact, in developed countries, judicial and governmental systems are efficient and already ensure the proper functioning of the market for promises. Conversely, in less-developed countries with weaker judicial systems, blockchain-based finance is likely to replace traditional finance. In the following sections, we will analyze some of the challenges and issues that blockchain-based finance must face before becoming a serious competitor to traditional finance, while simultaneously attempting to imagine possible future scenarios.

The rest of the paper is organized as follows. Section \ref{sec:fintech_regulation} discusses blockchain finance and regulation; Section \ref{sec:challenges_scenarios} explores major challenges including illegal activities, sustainability, market efficiency, and NFTs as digital property rights. Section \ref{sec:conclusions} concludes.

\section{Fintech and regulation}\label{sec:fintech_regulation}

As we have seen, Fintech relies on an entirely new technology to safeguard the market for promises. While traditional finance depends on the state's monopoly over the use of force and the proper functioning of the rule of law, Fintech relies on decentralized verification of promises by miners and their automatic execution through smart contracts. For these reasons, supporters of Fintech have historically been—and still are---``allergic'' to regulation, perceived as unnecessary, ineffective, and as an attempt by the state to curb a dangerous competitor. It is no coincidence that some of today's most prominent Fintech companies are legally headquartered in countries with less strict or less effective regulation. For example, Binance, the largest cryptocurrency exchange platform, is apparently registered in Malta despite operating in more than 100 countries worldwide. Similarly, FTX—recently in the headlines for a spectacular and painful default affecting hundreds of thousands of investors—was registered in the Bahamas, even though it sought to project the image of a responsible Fintech company through high-profile sponsorships (such as the FTX Arena in Miami, home to the NBA team Miami Heat), charitable donations (through the FTX Foundation), and good relations with U.S. regulatory authorities.

According to some observers, regulation would be entirely ineffective within Fintech due to the ease with which national borders and controls can be circumvented. Indeed, although the blockchain is transparent by nature, many transactions occur between so-called ``wallets” (a type of encrypted account), which are not easily traceable to the individuals or companies controlling them. A blockchain wallet is a tool that stores the public and/or private keys required to complete transactions on a blockchain. In this sense, wallets can be considered analogous to a deposit account at a bank. According to these observers, there would therefore be no viable alternative to blockchain self-regulation (see, e.g., the discussion in \cite{makarov2021blockchain}).

\begin{figure}[hbt!]
\hspace{0cm}
\centering
\includegraphics[scale=0.7]{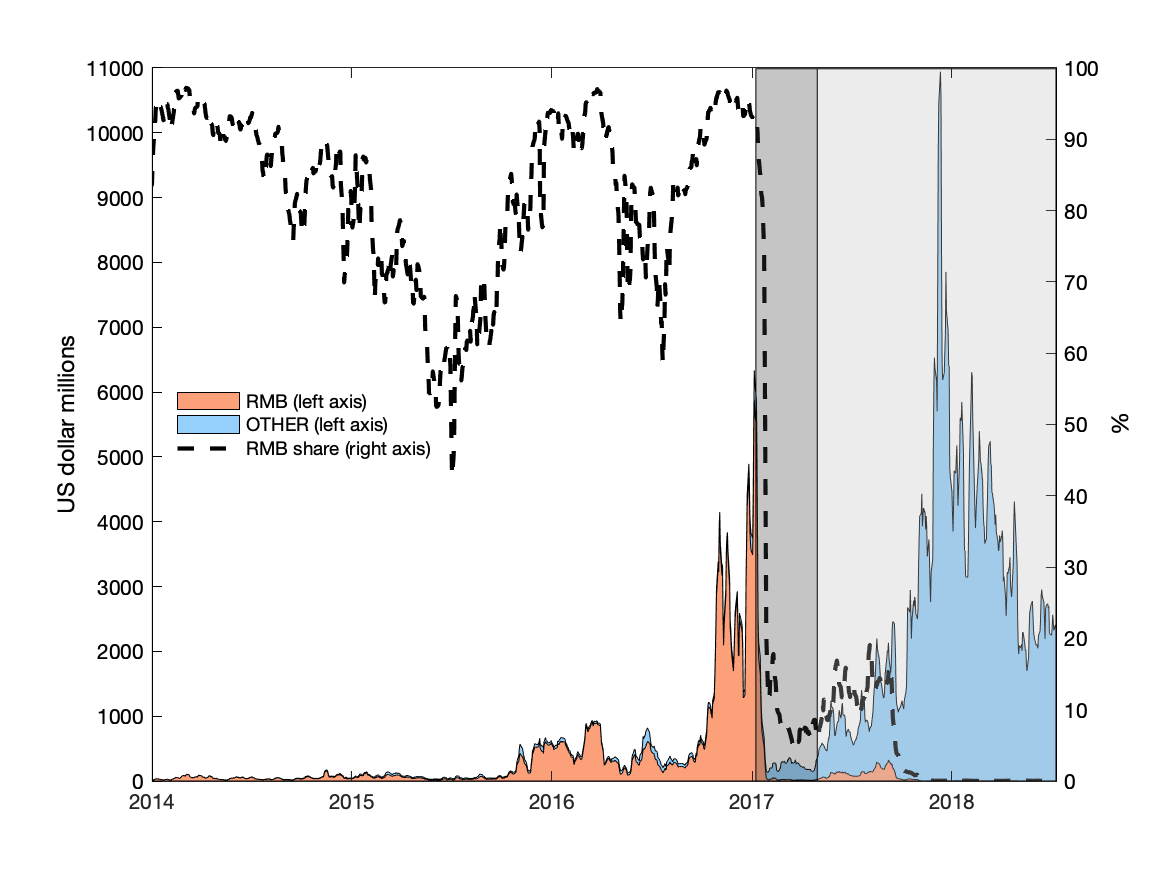}
\caption{This figure from \cite{borri2020regulation} illustrates the daily trading volume of Bitcoin denominated in RMB (light red) and all other fiat currencies combined (light blue), measured in millions of U.S. dollars (left axis). Data are aggregated from various cryptocurrency exchanges and currency pairs, spanning January 1, 2014, to July 31, 2018, and smoothed using a 5-day moving average. The black dashed line represents the share of Bitcoin trading volume in RMB (right axis). The dark grey shaded area denotes the first phase of the ``China shock'' (January 3 to April 27, 2017), while the light grey shaded area indicates the second phase of the ``China shock,'' extending to the end of the sample period. The entire shaded area represents the ``permanent shock,'' lasting from January 3, 2017, through July 31, 2018.}
\label{fig:borri_regulation}
\end{figure}

Empirical analysis, however, portrays a more complex and nuanced reality. For example, \cite{borri2020regulation} examine one of the major regulatory shocks that impacted the cryptocurrency world and analyze its effects. Their analysis shows significant negative impacts on both trading volumes and prices. Specifically, \cite{borri2020regulation} study a series of events that took place in China at the beginning of 2017, involving both a \textit{de jure} change in cryptocurrency regulations and a restrictive \textit{de facto} shift in the attitudes of Chinese authorities toward cryptocurrencies. The authors define these events collectively as the ``China shock'' due to their disruptive effects on trading volumes in China. As illustrated in Figure \ref{fig:borri_regulation}, within just a few days, the trading volume of Bitcoin denominated in RMB—the Chinese currency—drops from about 90\% of global volume to less than 10\%, eventually approaching zero by early 2018. Thus, the Chinese example demonstrates how public authorities can have disruptive effects on blockchain-based financial markets, at least within their directly controlled political region. However, the same figure also highlights a clear substitution effect. Following the regulatory shock, investors began trading Bitcoin in other currencies, such as the Yen. This latter effect partially arose because certain Fintech activities (for example, mining and trading operations) moved to other countries, and partially because Chinese investors started employing mechanisms to bypass controls, such as the use of VPNs (virtual private networks), which were already widespread in China to access other services unavailable due to public regulation, such as Google and Facebook. According to some observers, these same services were also used by U.S. users to access derivative products offered by FTX International, which were unavailable to them due to American regulations. Before its collapse, U.S. investors had access to FTX US, an entity separate from FTX International. Only the latter offered derivative products and margin lending. Initial data, however, suggest that many U.S. investors used VPNs to access services offered by FTX International. Similarly, a U.S. or British investor cannot access services provided by Binance. In contrast, an Italian investor can access Binance's services, except for derivatives and margin lending. It is noteworthy that most of these provisions are autonomously set by the platforms themselves, change frequently, and do not depend on market regulatory authorities.

Indeed, one of the main findings of \cite{borri2020regulation} is precisely that regulatory shocks in cryptocurrency markets typically generate significant externalities. For instance, both trading volume and demand for Bitcoin increased in neighboring countries following the ``China shock,'' particularly in Korea and Japan. Furthermore, the effects of this shock proved persistent, as evidenced by the subsequent relocation of mining activities—previously concentrated in China—to other countries, such as the United States. To this extent, the Cambridge Bitcoin Electricity Consumption Index (\href{https://ccaf.io/cbeci/mining_map}{https://ccaf.io/cbeci/mining\_map}) highlights the geographical redistribution of Bitcoin mining activity. Naturally, while causality is more clearly identifiable around the timing of regulatory shocks, many other variables may change over the longer term, and thus any causal conclusions should be interpreted with great caution.

The analysis described here suggests, on the one hand, that effective regulation of blockchain-based finance cannot be purely ``national'' but must instead rely on internationally coordinated agreements, not unlike those established in traditional finance. For example, consider the MiFID II directive in the European Union, which ultimately has ``global'' effects due to both the size of the European economy and the licensing requirements imposed on non-European operators seeking access to European savings and financial markets. On the other hand, the Chinese example illustrates that even when faced with a state possessing substantial coercive power, blockchain finance inherently facilitates the circumvention of potential barriers, such as through the use of software that masks IP addresses.

Other studies, considering a broader set of regulatory shocks, confirm the findings described above. For instance, \cite{auer2018regulating} construct a database containing numerous cryptocurrency regulation shocks and find that both trading volumes and valuations are significantly impacted. \cite{copestake2023crypto} extend the database of \cite{auer2018regulating} by including news regarding central banks' positions on introducing central bank digital currencies (CBDC). They find that trading volumes decline by up to 72\% following highly negative shocks, such as China's 2017 regulatory event, whereas shocks related to CBDCs have significant but considerably smaller effects.

Recent failures of major Fintech institutions—such as Celsius Network, specialized in crypto-lending; Three Arrows Capital, one of the largest crypto hedge funds; and notably, FTX, one of the largest cryptocurrency trading platforms—are likely to trigger a regulatory revolution in cryptocurrencies and blockchain-based finance. Indeed, particularly the failure of FTX has directly impacted hundreds of thousands of investors worldwide who were unable to properly evaluate the risks they faced. In the FTX case, users only discovered ex post that the trading platform probably also functioned as a bank, operating with leverage of about 10x by improperly lending (if the allegations are confirmed) the capital of unwitting users (incidentally, these loans appear to have been directed toward entities linked to FTX itself). All of this was possible because cryptocurrency trading currently occurs off the blockchain, as the speed required by investors is incompatible with the slower pace of blockchain information recording. In practice, investors typically deposit their crypto-assets into virtual wallets managed by exchanges. In return, the exchange provides an IOU, which exposes clients to counterparty risk, even though investors' holdings are meant to be segregated from the exchange’s funds, with the exchange acting merely as a custodian. Despite the security benefits, relatively few investors store their cryptocurrency assets in so-called cold wallets, which are physically isolated from the internet. Instead, most investors rely on custody services offered by exchanges. Decentralized exchanges, which offer on-chain trading, try to address these problems, although at the moment account only for a small fraction of trading in cryptocurrency markets (see, e.g., \cite{lehar2021decentralized} for details on the functioning of decentralized exchanges).

These events will undoubtedly lead to further restrictions on accessing certain services in specific markets. For instance, current regulations already effectively prevent (or strongly restrict) British citizens from crypto trading, and U.S. citizens from trading on platforms not registered in the United States.\footnote{Note that cryptocurrency regulation in the U.S. is expected to undergo significant changes during Donald Trump's second presidential term. At the time of writing, these regulatory changes have not yet been implemented.} Consequently, it is likely that the market share of more traditional Fintech operators, such as Coinbase, will increase at the expense of less transparent platforms like Binance, unless the latter significantly enhance their transparency. However, only regulation with global reach can be truly effective. Otherwise, opaque operators will retain incentives to offer services at lower costs, given that barriers can easily be bypassed using tools like the VPN software previously mentioned.

For this reason, proposals such as that of \cite{auer2022embedded}, which envision a new type of regulation more suitable for blockchain-based finance, are particularly interesting. Specifically, \cite{auer2022embedded} proposes the concept of ``embedded supervision,'' a regulatory system that automatically obtains information directly from the blockchain, thus reducing the need to collect and verify information manually. At present, it is challenging to determine the optimal regulatory solution for Fintech, but it is plausible that any effective solution will aim to leverage the functionalities and information provided by blockchain technology.

Stricter regulation of crypto and Fintech markets is not necessarily bad news for proponents of these financial innovations. Indeed, improved investor protection accompanied by greater transparency would likely attract more traditional savers toward Fintech products. Conversely, attempts at self-regulation within the Fintech sector seem likely—at least following recent events such as the collapse of FTX—to suffer a setback until new innovations enable greater transparency and investor oversight.

\section{Challenges and future scenarios for Fintech}\label{sec:challenges_scenarios}

In this section, we discuss some of the primary challenges and future scenarios for Fintech. Specifically, Section \ref{subsec:illegal} presents recent empirical evidence on the use of cryptocurrencies for illegal activities. Section \ref{subsec:sustainability} discusses the environmental sustainability of blockchain technology, fundamental to Fintech, from both current and future perspectives. Section \ref{subsec:trading_arbitrage} examines evidence regarding the efficiency of cryptocurrency markets, particularly price discrepancies across different markets and currency pairs, along with their implications for Fintech, especially concerning derivatives markets. Finally, Section \ref{subsec:NFT} introduces non-fungible tokens (NFTs) and highlights their key role in metaverse finance as digital property rights.

\subsection{Fintech and illegal activity}\label{subsec:illegal}

In the previous section, we discussed blockchain-based financial market regulation primarily from the perspective of investor protection. According to this viewpoint, regulation should aim at reducing risk---for instance, by limiting leverage---enhancing transparency, such as by publicly disclosing crypto-platform balance sheets and reserves, and maintaining a clear separation between investors' assets and those belonging to the platforms themselves.

A second reason prompting many observers to advocate for stricter regulation is related to the argument that the primary value of cryptocurrencies lies in their use as a medium of exchange for illegal activities. This argument is supported by the observation that cryptocurrencies generally lack fundamental value due to their nature as fiat currencies, even though privately issued and thus not backed by a central bank. This argument has received significant academic support in the work of \cite{foley2019sex}. Their research, among the first to analyze the use of cryptocurrencies in illegal activities, concludes that approximately one-fourth of Bitcoin users engage in illegal activities, and nearly half of Bitcoin transactions (or about \$76 billion annually) are related to illegal activities. Furthermore, although the use of Bitcoin for illegal purposes appears to have declined over time, the authors emphasize that this might simply reflect substitution towards other, more opaque cryptocurrencies. The results of this research imply that cryptocurrencies might add little value beyond facilitating activities that, as a society, we would like to discourage.

Firstly, it is important to note that the estimates provided by \cite{foley2019sex} are not unanimously accepted in the literature. For example, in another recent and highly influential study, \cite{makarov2021blockchain} present significantly different figures, criticizing the estimation methodology used by \cite{foley2019sex}. Indeed, \cite{makarov2021blockchain} explain that most Bitcoin transactions on the blockchain are related to trading activities and flows between trading platforms. These flows are structural, as they are necessary to align cryptocurrency prices across various trading platforms. According to their findings, illegal activities account for less than 3\% of total trading volumes. The radically different estimates ultimately stem from a distinct and more granular identification of blockchain participants achieved through machine learning and clustering techniques, as well as through the filtering of transactions deemed spurious.

Secondly, it is also worth emphasizing that numerous Fintech activities, such as the previously discussed examples of Uniswap and Aave, do not involve the exchange of Bitcoin but are instead based on the Ethereum blockchain. In a certain sense, it is overly simplistic to judge Fintech solely by looking at the current use of cryptocurrencies, especially Bitcoin, which is the most widely traded. Indeed, while Bitcoin and its blockchain remain, at least for now, primarily a currency or financial asset, Ethereum and its blockchain (alongside other blockchains, such as Solana or Algorand, for example) serve as breeding grounds for financial activities based on smart contracts. The Bitcoin blockchain was designed specifically to record transactions. For this reason, its blocks contain little information beyond the identifiers of the two parties involved in a transaction and the quantity of Bitcoin exchanged. Conversely, the Ethereum blockchain was designed to execute pieces of code, even complex ones. For this reason, the Ethereum blockchain has gradually become the foundation upon which various financial instruments and products have been developed. Ether (Ethereum’s native currency) serves as the medium of exchange on the Ethereum blockchain and is the currency with which miners are compensated. Consequently, the fundamental value of Ether depends on the value of the activities and applications built on the Ethereum blockchain.

Therefore, it is possible, if not probable, that blockchain technology will lay the foundations for finance that, in the near future, might appear more conventional (in the sense of more traditional), yet simultaneously fulfill the financial needs of our economy more efficiently. By analogy, at the end of the 1990s, few could have imagined that Amazon, initially an online bookseller, would become the giant we know today, with 75\% of its operating profits coming from cloud computing services (namely, Amazon Web Services, or AWS). A ``conventional'' Fintech finance is thus a financial system offering services to a much broader audience and partially or fully replacing traditional finance. Naturally, existing traditional financial operators could also provide services based on new technologies. Therefore, it is by no means inevitable that these traditional financial operators must disappear in the finance of the future.

\subsection{Fintech and sustainability}\label{subsec:sustainability}

Following the challenges of regulation and transparency regarding their use in illegal activities examined in previous sections, one of the further and most significant challenges facing Fintech is related to environmental sustainability. Indeed, according to some critics of blockchain-based finance, the decentralized verification of transactions through validation by miners is neither sustainable nor energy-efficient from an environmental perspective. Essentially, the amount of energy required to operate the various nodes comprising the blockchain would be so substantial that this technology becomes economically impractical and, above all, environmentally unsustainable. For example, frequent reference is made to the Bitcoin blockchain, which reportedly consumes approximately 200 TWh of electricity annually, an amount equivalent to the annual energy consumption of a country like Thailand. According to these observers, blockchain technology is destined to fail precisely because it is based on an unsustainable paradigm. In other words, the goal of greater security and independence compared to traditional exchange systems (such as information-exchange networks between banks and financial institutions) would be incompatible with environmental objectives.

However, this criticism does not necessarily represent a death sentence for blockchain technology. First of all, the history of innovation shows that technologies and products can become more energy-efficient and environmentally friendly over time. For instance, consider the enormous difference in emission levels between a Benz Patent Motor Car—one of the first automobiles with an internal combustion engine introduced in 1886—and a modern electric vehicle, such as a Tesla. Indeed, some of the latest technological innovations related to blockchain aim precisely to reduce pollutant emissions. The most notable case study is the Ethereum blockchain, which, in 2022, transitioned—via the so-called ``The Merge''—to a proof-of-stake validation system, moving away from the vastly more polluting proof-of-work system.\footnote{Proof-of-work is a system based on a competitive validation mechanism involving all the nodes within the blockchain. In contrast, proof-of-stake employs a subset of nodes, typically selected at random, to perform validation. The Ethereum Foundation, which manages the Ethereum blockchain, estimates that the proof-of-stake mechanism consumes approximately 99.95\% less energy compared to a proof-of-work system (see, for example, \href{https://content.ftserussell.com/sites/default/files/education_proof_of_stake_paper_v6_0.pdf}{https://content.ftserussell.com/sites/default/files/education\_proof\_of\_stake\_paper\_v6\_0.pdf}). However, according to some observers, the smaller number of nodes involved in the validation process in proof-of-stake mechanisms makes them inherently less secure than proof-of-work mechanisms.
} Although critics highlight that proof-of-stake validation mechanisms are less secure than proof-of-work mechanisms, it is not difficult to envision that technological innovation will continue, further improving security while simultaneously reducing environmental impact. A virtuous example in this regard is the Algorand blockchain, also based on proof-of-stake, which, since its inception, has made a firm commitment to environmental sustainability through partnerships with Climate Trade, a company specialized in offsetting the carbon footprint of businesses.\footnote{See, e.g., \href{https://www.algorand.foundation/impact-sustainability}{https://www.algorand.foundation/impact-sustainability}}

In conclusion, although the environmental challenge is significant and motivated by valid observations regarding the current energy consumption of certain blockchains, we can state that technological innovation is already oriented toward increased environmental sustainability, such that we can envision a greener blockchain in the future. Furthermore, the recent sharp increase in energy costs related to the war in Ukraine and tensions with Russia is providing an additional incentive to reduce energy consumption.

\subsection{Trading and arbitrage on cryptocurrency markets}\label{subsec:trading_arbitrage}

One characteristic of crypto markets, which is not always well understood, is that cryptocurrencies are traded simultaneously across different markets and against various other currencies—both fiat currencies, such as the dollar, and cryptocurrencies, such as Ethereum. For instance, one can buy Bitcoin using dollars on Coinbase, an exchange headquartered in the United States. Alternatively, one could purchase Bitcoin using dollars or euros on Kraken, another exchange that is very popular in Europe. Additionally, Bitcoin can be bought using Ethereum on Binance, an exchange where the primary transactions involve exchanging crypto for crypto. Consequently, there is no single price for Bitcoin but rather numerous prices depending on the exchange and/or the currency against which Bitcoin is traded.

In an efficient market without arbitrage limitations, the price of one Bitcoin should be the same across every market and currency. A sort of triangular arbitrage condition should hold and ensure that the\textit{ law of one price} is satisfied—meaning the same asset would have the same price in every market in the absence of transaction costs, once denominated in the same currency.

\begin{figure}[hbt!]
\hspace{0cm}
\centering
\includegraphics[scale=0.7]{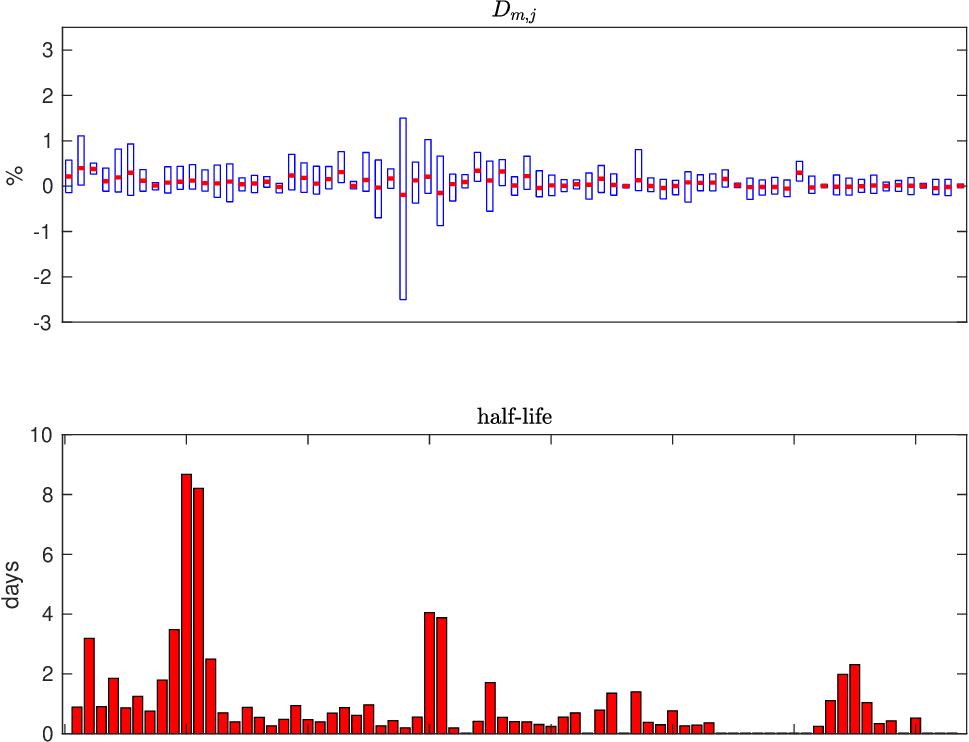}
\caption{This figure from \cite{borri2022cross} illustrates the magnitude and persistence of bitcoin discounts. The top panel of the figure presents a boxplot of bitcoin discounts ($D_{m,j}$) for all the pairs analyzed in the paper. For each box, the central red mark indicates the median discount, and the bottom and top edge of the box indicate the 25th and 75th percentiles. The bottom panel presents a bar plot where each bar corresponds to the half-life of one of the pairs in the baseline sample. To measure the half-life, \cite{borri2022cross} run an auto-regressive process of order 1 on discounts: the half-life is equal to $log(0.5)/ log(\rho)$, where $\rho$ is the persistence parameter, and captures the time it takes for a shock to dissipate by 50 percent. Note that standard augmented Dickey-Fuller tests reject the null of unit root for all pairs at standard significance levels. Data are daily from Cryptocompare and Thomson Reuters for the period 5/26/2015-5/25/2021.}
\label{fig:borri_discounts}
\end{figure}

The significant flows of Bitcoin recorded on the blockchain, which \cite{makarov2021blockchain} attribute to transfers between various exchanges, indicate the pressures from arbitrageurs aimed at maintaining parity among cryptocurrencies. However, as documented by \cite{makarov2020trading} and \cite{borri2022cross,borri2023cryptomarket}, substantial price differences exist for Bitcoin across different markets and against other currencies. For example, Figure \ref{fig:borri_discounts} illustrates the so-called ``Bitcoin discounts," or price discrepancies for Bitcoin in various markets. These differences are often large and volatile (upper panel) as well as persistent (lower panel).

The existence of Bitcoin (or crypto) discounts is problematic for at least two reasons. First, it may indicate inefficiencies in the crypto markets, potentially related to high transaction costs or other constraints on arbitrage activities. This is the argument advanced by \cite{makarov2020trading} and \cite{borri2023cryptomarket}, applicable especially to the largest discrepancies and prices observed in specific markets—particularly in closed economies or those with restrictions on international capital flows. Market inefficiencies would naturally represent an obstacle for Fintech, which relies heavily on cryptocurrencies and the markets where these discounts are observed. However, \cite{borri2023cryptomarket} also show that price discrepancies are significantly reduced in crypto-to-crypto transactions (e.g., Bitcoin for Ethereum). This finding suggests that cryptocurrencies can overcome restrictions imposed by capital controls when bypassing traditional fiat currencies. In such scenarios, crypto markets prove to be much more efficient, with smaller and far less persistent discounts.

The widespread adoption of stablecoins---cryptocurrencies regarded as the primary vehicle for many crypto transactions---suggests that future crypto markets may exhibit greater efficiency compared to the pre-2019 period. Stablecoins are cryptocurrencies whose value is pegged to a fiat currency, such as the dollar (or a basket of fiat currencies). Thus, stablecoins aim to combine the stability of traditional currencies with the unique features of cryptocurrencies. For this reason, ensuring maximum transparency for stablecoins and for the assets used to anchor their value to fiat currencies, such as the dollar, is crucial. There is indeed a risk that these assets may be illiquid or risky, leading to an imbalance between assets and liabilities and potentially causing a bank run. A recent prominent example of failure in this regard is the Terra-Luna stablecoin, whose value was algorithmically linked to another cryptocurrency (see, for details, \cite{liu2023anatomy}). Additionally, concerns—whether justified or speculative—have repeatedly emerged regarding the value and liquidity of the assets backing Tether, the largest stablecoin.

\cite{borri2022cross} demonstrate that many price discrepancies observed for Bitcoin across various markets are not necessarily indicators of mispricing or arbitrage opportunities (and therefore market inefficiency). Instead, \cite{borri2022cross} show that these discrepancies reflect the risks faced by potential arbitrageurs. The recent collapses of investment funds such as Three Arrows Capital and subsequently Alameda (the hedge fund affiliated with FTX), both of which attempted to profit from perceived mispricings, further support the argument that these strategies involve significant risk rather than pure arbitrage.

A second reason why price discrepancies for the same currency, such as Bitcoin, across different markets pose a challenge for Fintech is related to the potential development of crypto derivative finance. Consider a futures contract dependent on Bitcoin’s price: which price determines the futures' value if there are multiple prices for the same asset? This issue extends beyond futures and options contracts; credit activities also depend on the value (or multiple values, if prices differ) of collateral assets. What happens if different valuations exist for the same collateral?

Currently, Fintech operators set reference index prices using weighted averages from prices on the most liquid and efficient platforms. However, this approach is a second-best solution and remains vulnerable to significant price divergences even among the most liquid platforms during market stress. Given that such price discrepancies (discounts) are persistent, there is a risk that some investors may exploit these differences at the expense of others.

Developing a derivatives market is essential for the growth and maturity of Fintech. Without such a market, institutional investors face considerable difficulties allocating substantial portions of their portfolios to crypto assets, as derivatives are crucial for effective risk management and hedging strategies. Additionally, another important—more traditional—dimension of market efficiency involves the capacity of market prices to accurately reflect all available information. This aspect is particularly relevant for Fintech and blockchain finance due to the unique characteristics of the information environment in these markets. \cite{liu2021accounting} provides an insightful analysis of how information disclosed on blockchains influences the cryptocurrency market, demonstrating significant market reactions following these disclosures.

\subsection{The value of digital property rights}\label{subsec:NFT}

An analysis of scenarios and challenges in Fintech finance would be incomplete without discussing one of the latest technological innovations related to blockchain: non-fungible tokens, or NFTs. NFTs are tokens—assets native to an existing blockchain (e.g., Ethereum blockchain). Unlike generic tokens, NFTs are unique and thus non-fungible. Essentially, NFTs are unique digital identifiers that cannot be copied or replaced and are registered on the blockchain. Consequently, ownership of an NFT can be transferred, allowing NFTs to be sold and exchanged.

Although NFTs are currently known primarily for representing digital items with eccentric names, such as the famous (and expensive) Bored Ape Yacht Club or Cryptokitties, many observers see enormous potential in NFTs for the future of finance and the economy—potentially even greater than that of cryptocurrencies. Indeed, property rights have always been fundamental to growth and technological innovation. In a sense, NFTs represent a step forward in the innovation process, laying the foundations for the concept of digital property rights within the digital economy, or more specifically, the economy of the metaverse.

Studying the NFT market presents researchers with several challenges. NFTs are traded across multiple platforms, such as OpenSea and Atomic, and as non-fungible assets, they resemble real assets like real estate more closely than financial assets such as cryptocurrencies. Indeed, similar to real estate, NFTs exhibit low liquidity, as individual NFTs are infrequently traded. \cite{borri2022economics} provide the first comprehensive economic and financial analysis of the NFT market. The authors compile a dataset containing information on nearly the entire universe of NFT transactions, encompassing tens of millions of transactions. They gather this data directly from trading platforms and subsequently validate it by cross-referencing with blockchain data. The collected dataset is extremely rich: in addition to transaction prices, it includes detailed information about the traded asset, associated royalties, identities of transaction participants, and allows for the construction of precise indicators of uniqueness. The authors employ this data in a hedonic regression analysis to identify determinants of NFT valuations. The results indicate that attributes such as ``rarity'' or royalties payable to the creator, along with various fixed effects capturing the creator and collection, explain a substantial portion of the price variability in the NFT market, akin to traditional asset markets.

Subsequently, the authors use this data to construct the first NFT market index based on the repeat-sales methodology (see, \cite{bailey1963regression}). This methodology extracts the common market component, such as those observed in real estate or art markets, by controlling for individual asset characteristics through price differences observed in successive sales of the same asset. This approach was popularized by the Case-Shiller index for the real estate market (see, \cite{shiller1991arithmetic}). \cite{borri2022economics} demonstrate that this is also the most appropriate methodology for the NFT market and utilize the constructed index to analyze market properties, such as concentration and its relationship to other markets trading illiquid assets, like art or OTC markets. For instance, the authors find significant segmentation in the NFT market. Examining the largest NFT investors, they observe portfolios with very low diversification, where the majority of NFTs are purchased from a limited number of sellers.

As previously explained, constructing an NFT market index is fundamental, representing the first step toward developing derivative financial instruments that allow investors to invest in the market and hedge their investments. Indeed, futures, options, and even simpler collective investment instruments are inconceivable in a market without a representative index. Hence, defining a market index is indispensable for the market's future development by facilitating the entry of venture capital from investors betting on the potential role NFTs may play in the metaverse. In a sense, creating an NFT market index enables investors to perform securitization, transforming an illiquid asset into a liquid one, similarly to how the Case-Shiller index contributed to enhancing liquidity in the U.S. real estate market.

Finally, \cite{borri2022economics} demonstrate how analyzing visual images of NFTs using machine learning techniques, such as neural networks, significantly enhances valuation methods, a core aspect of corporate finance. The authors adopt a methodology similar to that used by \cite{aubry2023biased}, who examine the impact of visual images on the valuation of artworks. However, unlike findings in the traditional art market, \cite{borri2022economics} show that incorporating visual analysis substantially improves the predictive accuracy of valuation models in the NFT market. This result holds broader implications, suggesting potential applications for valuation practices in other alternative asset classes where visual characteristics are critical (e.g., real estate, automobiles, etc.).

\section{Conclusions}\label{sec:conclusions}

This article has discussed the possible, if not likely, future scenarios and challenges facing Fintech. Blockchain is a technological innovation that holds the potential to radically transform our financial markets by providing an alternative approach to managing the promise market. However, to fully or partially replace traditional finance, it must overcome significant challenges such as regulation, environmental sustainability, associations with illegal activities, and improving the efficiency of cryptocurrency markets. For these reasons, the future of Fintech is likely to be more conventional but also more transparent, efficient, and regulated.


\newpage
\bibliographystyle{rfs}
\bibliography{bib_fintech}

\end{document}